# First order transition in two dimensional coulomb glass

**Preeti Bhandari[1] and Vikas Malik[2]**

1Department of Physics, Jamia Millia Islamia, New Delhi 110025, India
2Department of Physics and Material Science, Jaypee Institute of Information Technology, Uttar Pradesh, 201304, India

preetibhandari15@gmail.com

**Abstract.** We have studied the ground states of two dimensional lattice model of Coulomb Glass via Monte Carlo annealing. Our results show a possibility of existence of a critical disorder ($W_c$) below which the system is in the charge ordered phase and above it the system is in the disordered phase. We have used finite size scaling to calculate $W_c$ = 0.2413, the critical exponent of magnetization $\beta$ = 0 indicating discontinuity in magnetization and the critical exponent of correlation length $\nu$ = 1.0. The distribution of staggered magnetization for different disorder strengths shows a three peak structure. We thus predict that two dimensional Coulomb Glass shows a first order transition at T=0.

## 1. Introduction

Since the introduction of Random field Ising Model, a lot of discussion has been done on the possibility of existence of ferromagnetic ordering below a critical disorder strength ($W_c$) in two dimensional (2D) RFIM. From Imry-Ma arguments [1], one finds that the lower critical dimension ($d_c$) below which the ferromagnetic ordered phase gets destroyed is $d_c \leq 2$, where d = 2 was considered as a limiting case. By doing field theoretical calculations [2,3], later it was suggested that $d_c$ = 3. In 1983, Binder [4] gave an argument that the roughening of domain walls would lead to the destruction of ferromagnetic ordering in 2d RFIM. Then an exact result was given by Bricmont and Kupiainen [5] in 1987 where they showed that there exist a ferromagnetic ordering in three dimensional (3d) RFIM. A rigorous theoretical proof was then given by Aizenman and Wehr [6] in 1989 that no long range ordering is possible in 2d RFIM case. Scaling theory for 3d RFIM assuming second order transition at T=0 leads to modified hyperscaling laws [7].

However, in contradiction to all the above arguments Frontera and Vives [8] in 1999 showed numerical signs of transition in 2d RFIM at zero temperature below a critical random field strength. In 2012, Aizenmann [9] too claims existence of phase transition in 2d RFIM. Spasojevic et al [10] have also given some numerical evidences which proves the existence of phase transition in 2d non-equilibrium RFIM at T = 0. Recently, Suman and Mandal [11] have studied some aspects of non-equilibrium 2d RFIM at low temperature. They have commented that 2d RFIM exhibits a phase transition in the disorder parameter even at T > 0. These numerical work suggests a possibility of presence of long range ordering in 2d RFIM at low disorder strengths.

If we assume that the above arguments which are claiming that at T = 0 there is a finite disorder strength below which long range ordering in 2d RFIM exist, then the order of transition is another question which is not answered very clearly. This question remains unclear even for 3d RFIM where there is no controversy over the presence of phase transition at zero temperature. Some earlier work suggested a first order transition [12-18] but there are arguments [19-22] which favour second order transition in 3d RFIM. Since the value of magnetic component β/ν is very small, it is difficult to determine whether at transition the magnetization vanishes continuously or discontinuously. Numerical study of 3d RFIM at T = 0 claims transitions to be of continuous type [23-26] but the value of the critical exponent of magnetization (β) is very close to zero in all the papers.

We are here interested in determining whether a critical behaviour similar to 2d RFIM exist in the case of 2d Coulomb Glass (CG). Coulomb Glass is defined as a system where all electron states are localized and interactions takes place via long-range Coulomb potential. Since at low temperature, localized electrons are unable to screen the Coulomb interactions effectively, it will be interesting to see how differently these long range interactions effect the phase transition from the RFIM. In our previous work [27] we found the transition from COP to disordered phase was driven by rearrangement of domain wall of the metastable state in charge ordered phase (COP) as W was increased to give disordered phase which indicates phase coexistence around transition. In this paper we have calculated the critical exponents using finite size scaling analysis. We have also investigated the distribution of staggered magnetization which shows a three peak structure characterstic of a first order transition.

## 2. Model

We have considered lattice model of CG in 2d as defined by Davies et al [28]. The model is a regular lattice of sites where on each site electron states are localized. The Hamiltonian of this system, defined over a spin configuration $\{S_i = \pm 0.5\}$ is given as

$$H = \sum_i \phi_i S_i + \frac{1}{2} \sum_{i \neq j} J_{ij} S_i S_j \qquad (1)$$

where the random on-site energies $\varphi_i$ are chosen independently from a probability distribution $P(\varphi_i)$ which is considered as

$$P(\phi) = \begin{cases} \dfrac{1}{2W}, & if -W/2 \leq \phi \leq W/2 \\ 0, & otherwise \end{cases} \qquad (2)$$

The unscreened Coulomb interactions is defined as $J_{ij} = e^2/\kappa R_{ij}$ where κ is the dielectric constant and $R_{ij}$ is the distance between sites i and j which is calculated using periodic boundary conditions. Since the system under consideration possesses particle-hole symmetry hence the chemical potential μ = 0. We have measured energies in units of $e^2/\kappa a$ where a ≡ 1 is the lattice constant. We have only considered the case of half filling where the number of electrons are half of the total number of sites (N) in the system.

We assume that there exist a critical disorder $W_c$ below which phase is charge ordered and above which it is disordered [29]. Now for a CG system a COP implies Antiferromagnetic ordering as $J_{ij} > 0$. So staggered magnetization is the order parameter which is defined as

$$M_s = \left\langle \frac{1}{N} \sum_i \sigma_i \right\rangle \qquad (3)$$

where $\sigma_i = (-1)^i S_i$ and <...> is the ensemble average performed over different disorder configurations.

## 3. Numerical simulation

We have first used Monte Carlo (MC) annealing technique to reach the minimum possible energy state. The initial configuration used for the simulation was completely random with half sites assigned $S_i$ = 0.5 and the second half assigned with $S_i$ = -0.5 . The annealing was done using the Metropolis algorithm [30]. This algorithm constitutes of a random walk in the entire space of all possible configurations of the system. Since the number of electrons are conserved, one uses Kawasaki Dynamics. A single MC step is performed by randomly choosing two sites i and j, where one site is unoccupied ($S_i$ = 0.5) and the second site is occupied ($S_i$ = -0.5) for spin-exchange. This exchange is not always executed, which is possible in a simple MC procedure. If the exchange of spins corresponds to relaxation i.e. if the final state of the system yields an energy level which is lower than the initial state, then the exchange is done with a probability

$$P_{ij} = 1 \quad (4)$$

On the other hand if the exchange corresponds to a thermal excitation into a state whose energy is higher than the initial state, let us assume by an amount $\Delta E_{ij}$ then the spin-exchange is done with a probability given as

$$P_{ij} = \exp\left(\frac{-\Delta E_{ij}}{\kappa T}\right) \quad (5)$$

where $\Delta E_{ij} = \varepsilon_j - \varepsilon_i - 1/R_{ij}$ is calculated using the single particle Hartree energy ($\varepsilon_i$) [31] which is expressed as

$$\varepsilon_i = \phi_i + \sum_{i \neq j} J_{ij} S_j \quad (6)$$

The Hartree energy of the initial random system is calculated and stored. If the chosen site is the original position of the electron then the Hartree energy remains unchanged but if the electron hops to a new site, then we update all the Hartree energies of the system. When the system is in the low temperature state then the electron chooses its initial state site with a very high probability, so we do not have to recompute our Hartree energies again which speeds up our simulation considerably at low temperature regime. Annealing was done from $\beta = 1$ to $\beta = 100$ (where $\beta = 1/T$) for different system sizes (L=16,32,48). At low temperature we increased the MC steps and our longest run was $5 \times 10^5$ for each $\beta$ above 20. The investigation were carried out at different disorders starting with W=0.0 to W=0.50 for all L.

Once annealing was over we then identified the clusters which were created in the minimum energy state using the Hoshen-Kopelman algorithm [32]. The cluster with nearest-neighbour sites that have same $\sigma_i$ was defined as domain. After identifying the domains we then investigated each domain's interactions with all the other domains present in a single disorder realization. We did not find any significant domain-domain interactions present. The domains were then flipped one by one if it decreased the energy of the system. The final state in which no more flipping was possible was then considered as the ground state. By this method we do not claim that we obtained the exact ground state.

## 4. Results and discussion

After obtaining ground state, we investigated the average size of the largest and second largest domain at each disorder for the largest system size that we have considered (L=48). Figure 1 shows the volume of the largest and the second largest domains in the ground state. For W<0.20 all the domains got flipped and we found single domain picture which indicates a COP. But on further increasing the disorder strength the size of the second largest domains starts increasing indicative of occurrence of disordered phase. The picture is similar to 2d and 3d RFIM where the ordered state breaks into two large domains [33]. For each disorder strength (W) we have then recorded the average staggered magnetization per spin. The staggered magnetization shows sharp decrease with disorder as shown in figure 2. Calculation was done for different system sizes, L=16, 32, 48 and disorder averaging was done over 800, 400 and 350 realizations of random field respectively.

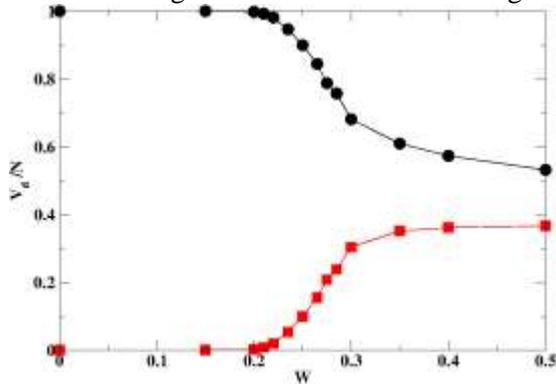 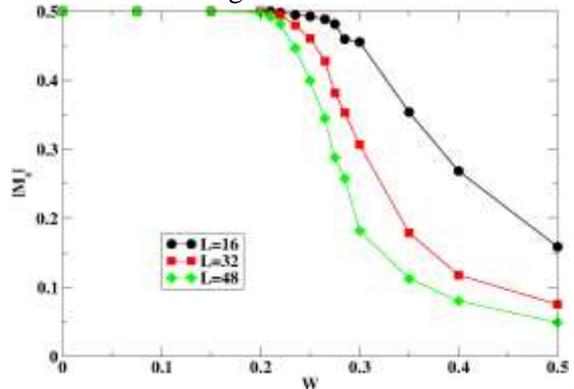

**Figure 1.** Largest (●) and second largest (■) domain size ($V_d$) divided by the system size N = $L^2$ at different disorder strength (W) for L=48

**Figure 2.** Behavior of staggered magnetization vs disorder strength W for system size L=16, 32, 48.

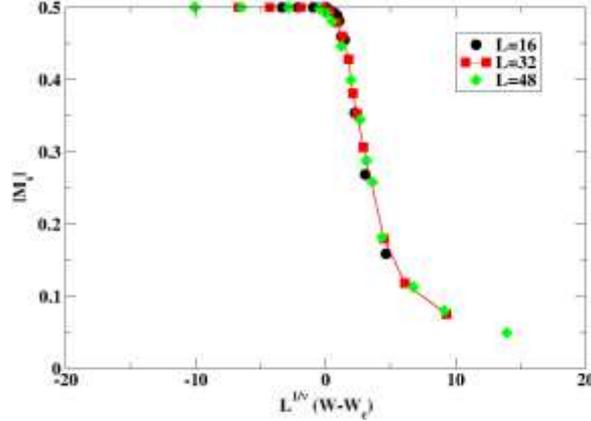

**Figure 3.** Scaling plot of the staggered magnetization $|M_s|$ with the parameters for $W_c = 0.2413$ and $1/\nu = 1.0$. The full line connects the points for L=32 as a guide for the eyes. Here the data for $|M_s|$ are not rescaled by the factor $L^{-\beta/\nu}$, which implies that $\beta = 0$.

To find the exact value of critical disorder ($W_c$) and to determine the order of transition we have used standard finite size scaling function for $M_s$ which is defined as

$$M_s = L^{-\beta/\nu}\tilde{M}_s((W-W_c)L^{1/\nu}) \qquad (7)$$

where $\beta$ is the exponent of staggered magnetization and $\nu$ is defined as correlation length exponent. The scaled data is shown in figure 3, where $W_c = 0.2413$; $\beta = 0$; and $1/\nu = 1.0$. The value of $\beta = 0$ could be indicative of a first order transition but as mentioned in the introduction is not a conclusive proof for a first order transition.

In our previous paper [27] we studied the system near the transition region where we found that the domain wall of the metastable state in the charge ordered phase (COP) shifts to give disordered ground state. This indicates that the transition is first order type as the free energy which is equal to the energy at zero temperature has three minimas which are centered around $M_s = \pm 0.5, 0$. This also shows that for each disorder configuration the staggered magnetization changes discontinuously at $W_c$. Such discontinous jumps for bond energy were also reported for 3d RFIM [18, 22, 34]. The domains of the metastable state in the COP and the domains in ground state of disordered phase are plotted in figure 4 for L=48.

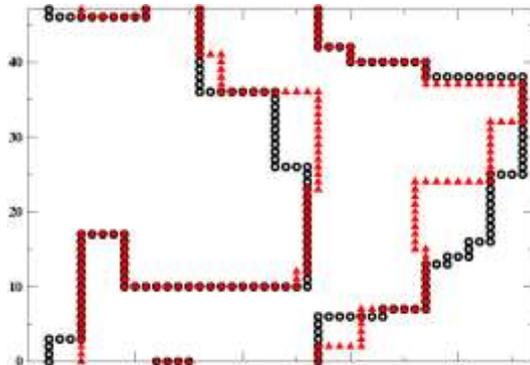

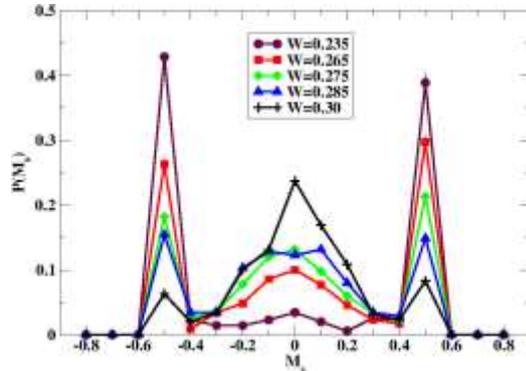

**Figure 4.** Domain wall of metastable state at W=0.235 and domain wall of ground state at W=0.25 are plotted for one of the configurations at L=48.

**Figure 5.** Histogram of the staggered magnetization distribution for L=48 system for different W's.

We have then investigated the distribution of statggered magnetization at different W for L=48. As shown in figure 5, for W<$W_c$, peaks are centered around $M_s = \pm 0.5$ and for W>$W_c$ around $M_s = 0$ as expected. In the transition region $0.265 < W < 0.285$, one gets a three peak structure. The transition region is same as one gets in the staggered magnetization vs W graph in figure 2 for L=48. This three peak picture around the transition region is a further indication of coexisting phases which is a characteristic feature of first order transition. A similar picture for 4d bimodal RFIM was reported by Swift et al [35] but no coexisting phases were found for 3d bimodal and Gaussian RFIM.

## 5. Conclusions

To summarize we have investigated the ground state of 2d lattice model of Coulomb Glass. Using finite size scaling analysis, we have shown existence of critical disorder below which the phase is ordered (antiferromagnetic ordering) and above $W_c$ = 0.2413 it is disordered. The critical exponent of magnetization, β = 0 indicates that the transition is first order. To verify this point further we have also shown the presence of coexisting phases in two ways. Firstly, the domain in the metastable state of the COP shifts to give ground state in the disordered phase and secondly the three peak picture in the distribution of $M_s$ around the transition region characteristic feature of first order transition.

**References**
[1]  Imry Y and Ma S K 1975 *Phys. Rev. Lett.* **35** 1399
[2]  Young A P 1977 *J. Phys. C* **10** L257
[3]  Parisi G and Sourlas N 1979 *Phys. Rev. Lett.* **43** 744
[4]  Binder K 1983 *Z Phys. B* **50** 343
[5]  Bricmont J and Kupiainen A 1987 *Phys. Rev. Lett.* **59** 1829
[6]  Aizenman M and Wehr J 1989 *Phys. Rev. Lett.* **62** 2503
[7]  Bray A J and Moore M A 1985 *J. Phys. C* **18** L927
[8]  Frontera C and Vives E 1999 *Phys. Rev. E* **59** R1295
[9]  http://ricerca.mat.uniroma3.it/ipparco/convegno70/seminars.html
[10] Spasojevic D S Janicevic and Knezevic M 2011 *Phys. Rev. Lett.* **106** 175701
[11] Sinha S and Mandal P K 2013 *Phys. Rev. E* **87** 022121
[12] Houghton A, Khurana A and Seco F J 1985 *Phys. Rev. Lett.* **55,** 856
[13] Young A P and Nauenberg M 1985 *Phys. Rev. Lett.* **54** 2429
[14] Brezin E and De Dominicis C 1998 *Europhys. Lett.* **44** 13
[15] Machta J, Newman M E J and Chayes L B 2000 *Phys. Rev. E* **62,** 8782
[16] Fytas N G and Malakis A 2008 *Eur. Phys. J. B* **61** 111
[17] Fytas N G and M-Mayor V 2013 *Phys. Rev. Lett* **111** 019903
[18] Wu Y and Machta J 2006 *Phys. Rev. B* **74** 064418
[19] Villian J 1985 *Phys. (Paris)* **46** 1843
[20] Fisher D S 1986 *Phys. Rev. Lett.* **56** 416
[21] Natterman T 1997 *Spin Glasses and Random Fields*, ed Young A P (Singapore: World Scientific)
[22] Hartmann A K and Young A P 2001 *Phys. Rev. B* **64** 214419
[23] Middleton A A and Fisher D S 2002 Phys. Rev. B **65** 134411
[24] Dukovski I and Machta J 2003 *Phys. Rev. B* **67** 014413
[25] Ogielski 1986 *Phys. Rev. Lett.* **57**, 1251
[26] Vink R L C, Fischer T and Binder K 2010 *Phys. Rev E* **82** 051134
[27] Bhandari P and Malik V 2016 *arxiv:1610.06750v1* [cond-mat.dis-nn]
[28] Davies J H, Lee P A and Rice T M 1984 *Phys. Rev. B* **29** 4260
[29] Goethe M and Palassini M 2009 *Phys. Rev. Lett.* **103** 045702
[30] Metropolish N, Rosenbluth A W, Rosenbluth M N, Teller A H and Teller E 1953 *J. Chem. Phys.* **21** 1087
[31] Efros A L and Shklovskii B I 1975 *J. Phys. C* **8**, L49
[32] Hoshen J and Kopelman R 1976 *Phys. Rev. B* **14,** 3428
[33] Esser J, Nowak U and Usadel K D 1997 *Phys. Rev. B* **55** 5866

[34]     Wu Y and Machta J 2005 *Phys. Rev. Lett* **95** 137208
[35]     Swift M R, Bray A J, Maritan A, Cieplak M and Banavar J R 1997 *Europhys. Lett.* **38** 273

**Acknowledgments**
We wish to thank NMEICT cloud service provided by BAADAL team, cloud computing platform, IIT Delhi for the computational facility. Preeti Bhandari acknowledges UGC, Govt. of India for financial support through UGC-BSR fellowship (F.25-1/2013-14(BSR)/7-93/2007(BSR)).